\documentclass[final]{naturejournal}

\newcommand{\vE}{\vec{E}}
\newcommand{\vB}{\vec{B}}
\newcommand{\vD}{\vec{D}}
\newcommand{\vH}{\vec{H}}
\newcommand{\vj}{\vec{j}}
\newcommand{\pos}{\vec{r}}
\newcommand{\vp}{\vec{p}}
\newcommand{\vk}{\vec{k}}
\newcommand{\vt}{\vec{t}}
\newcommand{\oM}{M}
\newcommand{\oN}{N}
\newcommand{\oV}{V}  
\newcommand{\Green}{G}  
\newcommand{\preinv}{\Gamma^{-1}}

\title{A physics-defined recurrent neural network to compute coherent light wave scattering on the millimetre scale}

\author{Laurynas Valantinas\orcid{0000-0002-0591-1407}}
\author{Tom Vettenburg\corresponding\orcid{0000-0002-5508-1130}}
\affil{School of Science and Engineering, University of Dundee, DD1\,4HN, UK}
\affil{\email{t.vettenburg@dundee.ac.uk}}

\begin{document}
    \begin{abstract} 
        \internallinenumbers
        Heterogeneous materials such as biological tissue scatter light in random, yet deterministic, ways. Wavefront shaping can reverse the effects of scattering to enable deep-tissue microscopy. Such methods require either invasive access to the internal field or the ability to numerically compute it.
        However, calculating the coherent field on a scale relevant to microscopy remains excessively demanding for consumer hardware.
        Here we show how a recurrent neural network can mirror Maxwell's equations without training. By harnessing public machine learning infrastructure, such \emph{Scattering Network} can compute the $\SI{633}{\nano\meter}$-wavelength light field throughout a $\SI{25}{\milli\meter^2}$ or $176^3\SI{}{\micro\meter^3}$ scattering volume. The elimination of the training phase cuts the calculation time to a minimum and, importantly, it ensures a fully deterministic solution, free of any training bias.                       
        The integration with an open-source electromagnetic solver enables any researcher with an internet connection to calculate complex light-scattering in volumes that are larger by two orders of magnitude.
    \end{abstract}
    
    \noindent
    Machine learning has brought automation to areas that were traditionally considered to be the exclusive remit of human intelligence. More recently it was found that artificial neural networks can also circumvent the curse of dimensionality in challenging scientific computations~\cite{Jumper21}. This brings once-intractable problems within reach of current machine learning infrastructure.
    
    Advances in optics and photonics increasingly rely on our ability to accurately compute how light propagates and scatters as dictated by Maxwell's Equations~\cite{Molesky18, Thendiyammal20, Chen20, Jo22, Xue22, Yamilov22}. Coherent wave calculations provide essential information that can be impractical or even impossible to obtain experimentally. Notwithstanding, solving Maxwell's equations in complex heterogeneous materials such as biological tissue remains a long-standing challenge in its own right. The status quo has recently been challenged by the Convergent Born Series Method~\cite{Osnabrugge16,Krueger17,Vettenburg22}, raising hopes that light-wave scattering calculations can be brought into the realm of microscopy. Still, a volume of relevance to microscopy could span hundreds of wavelengths per dimension, while most algorithms demand a large number of samples per wavelength to keep error accumulation and numerical dispersion in check~\cite{Taflove05}. A significant amount of computer memory is thus required to solve for the billions of free parameters, a number on par with that of OpenAI’s GTP-3 deep learning language model~\cite{Floridi20}.
    
    In what follows, we show how Maxwell's equations can be rephrased as a recurrent neural network. Its topology models multiple scattering within heterogeneous materials, hence we will here refer to it as a \emph{Scattering Network}. Instead of being physics-guided or physics-based\cite{Vamaraju19, Kellman19, Kellman20, Willcox21, dArco22, Tang22}, its connections are directly dictated by the laws of physics. This eliminates the need for computationally expensive training while ensuring that its response is fully deterministic. Its highly-parallel neural network structure allows us to effectively harness the economy of scale of cloud-based machine learning infrastructure. We demonstrate its potential by computing the light-scattering of a point source embedded deep within a millimetre-scale heterogeneous structure. Next, we show how the efficiency of cloud-based calculations makes it possible to calculate the complete scattering and deposition matrices~\cite{Bender22}. Finally, we analyse and compare the performance to the Convergent Born Series Method on a desktop, a GPU workstation, and on the cloud; respectively improving the efficiency $20$ and $279$-fold.

    \section{Results}
        \subsection{Maxwell's equations as a Scattering Network.}
            \begin{figure*}[p]
                \centering\includegraphics[width=\textwidth]{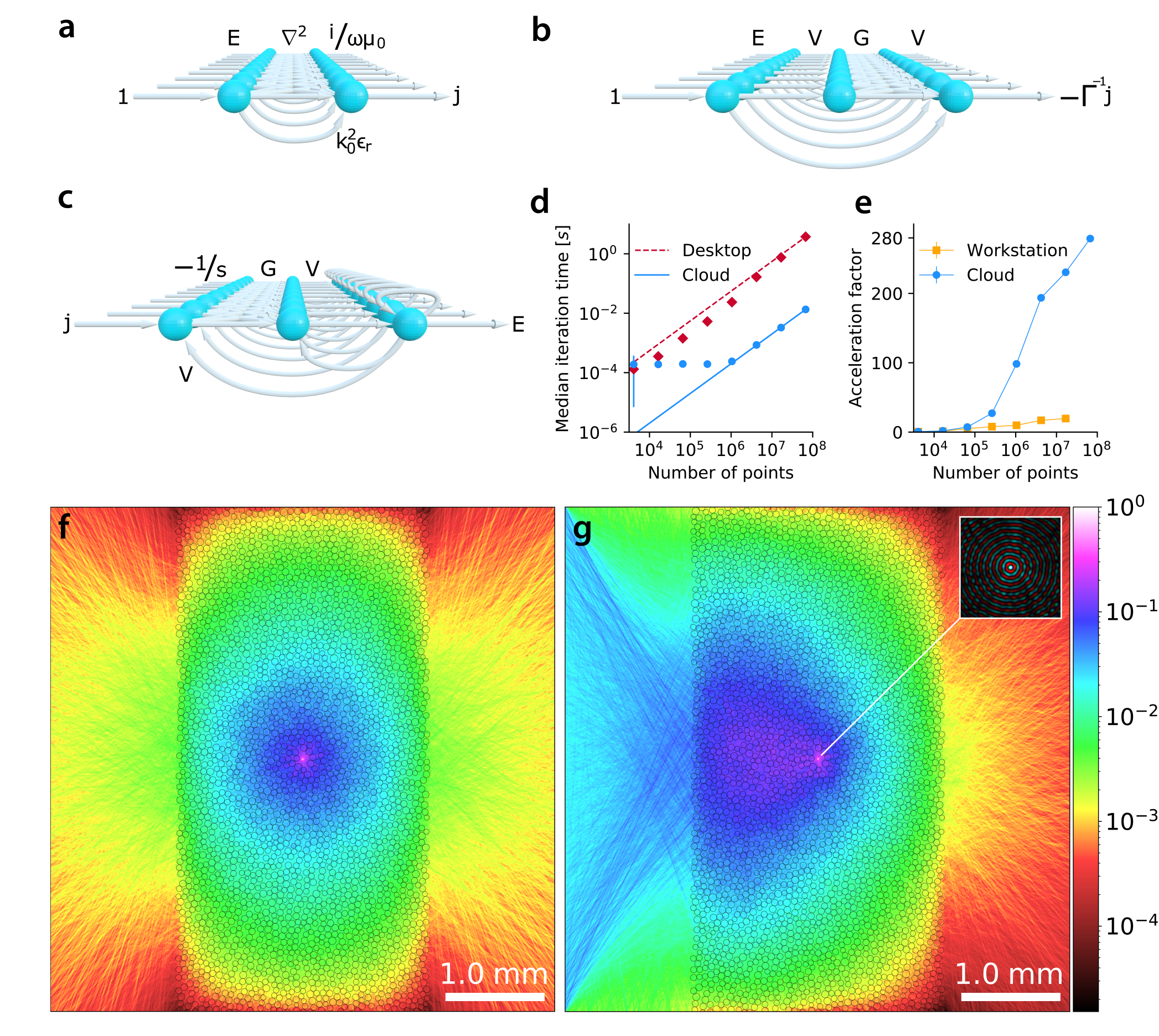}
                \internallinenumbers
                \caption{\textbf{The Scattering Network --- mapping wave scattering onto the structure of a neural network.} (\textbf{a}) Training a large neural network with this specific design yields the solution to Maxwell's equations. This network involves parallel paths and a convolutional layer to represent the Laplacian, $\nabla^2$. The spheres represent neurons that sum all inputs with a linear activation function and without bias. Each neuron corresponds to a location, $r$, in the discretised calculation space. Arrows represent connections with their multiplicative weights indicated above. Only the first (left-most) network layer, $E(r)$, requires training. When this network is presented with a unit input on the left, it returns a current distribution, $j(r)$, on the right that emits the field $E(r)$. When the network is trained for a given source, $j(r)$, the trained connection weights in the first layer correspond to the scattered field solution, $E(r)$. In spite of its simple topology, training this network was found to be challenging even for the smallest problems.
                (\textbf{b}) The network corresponding to the preconditioned system, $\Gamma^{-1}ME = \Gamma^{-1} j$. Although this topology is more complex, it ensures efficient and monotonic convergence of the training. The preconditioner, $\Gamma$, is a function of the modified permittivity, $\oV$, and the shifted Green's function, $\Green$. The out-of-plane unlabelled connection skips a layer of neurons, not unlike a residual block~\cite{He16}, though it has unity weights in this case. (\textbf{c}) The Scattering Network is recurrent. Its out-of-plane connections feed the final layer's output back, either identically, or multiplied by $\oV$ as indicated. Unlike the forward networks, the Scattering Network is fully defined by the physics. Without the need for training, it can efficiently and deterministically infer the solution field, $E(r)$. (\textbf{d,e}) Efficiency evaluation of the recurrent Scattering Network implemented using the machine learning library PyTorch~\cite{Paszke2019}. The wide availability of this framework facilitated both local and cloud deployment. We measured the median recurrence times of the PyTorch implementation on an NVidia Quadro P2000 GPU (Workstation) and on Google Colab (Cloud), for problems with sizes extending over four orders of magnitude. (\textbf{d}) The median recurrence time of the PyTorch implementation (Cloud), compared to that of the highly-efficient Convergent Born Series Method~\cite{Osnabrugge16} on an Intel i7 CPU (Desktop). The Scattering Network can be readily implemented using PyTorch for cloud deployment. It can be seen that, for all but the smallest problems, this reduces the calculation time by two orders of magnitude. The largest problems are solved $279\times$ faster. (\textbf{e}) The acceleration factor with respect to a desktop, of the recurrent Scattering Network on a GPU workstation and the cloud. (\textbf{f-g}) Scattering and refocussing of coherent waves in a heterogeneous material of size $\SI{5}{\milli\meter} \times \SI{5}{\milli\meter}$ or $8,000\;\lambda \times 8,000\;\lambda$ with vacuum wavelength $\lambda=\SI{633}{\nano\meter}$. The Scattering Network (Fig.~\ref{fig:intro}c) enabled the computation of each field within just $\SI{108}{\minute}$. (\textbf{f}) A heterogeneous material and the scattered wave emitted by a guide-star embedded deep within it. (\textbf{g}) The phase-conjugated field refocusing onto the target position from the left-hand side. Inset: the $100\times$-magnification shows how the field converges on the focal point. Brightness indicates amplitude and hue indicates phase (legend in Fig.~\ref{fig:sm_structure}e).}
                \label{fig:intro}
            \end{figure*}
        
        	We start by describing a two-layer forward neural network that mirrors the electromagnetic wave equation (Fig.~\ref{fig:intro}a). Next, we incorporate an extra layer for preconditioning to significantly accelerate training (Fig.~\ref{fig:intro}b). This preconditioned network is then transformed into the physics-defined recurrent Scattering Network that does not require training (Fig.~\ref{fig:intro}c).
     
            A large neural network model generally must be \emph{trained} to produce the target output values, $\vt$, for a large number of test cases. Training is the optimisation of a parameter vector, $\vp$, to minimise the difference between neural network's output and the training target. The optimal parameter vector is given by
            \begin{equation}\label{eq:neural_network_optimisation}
                \hat{\vp} = \argmin_{\vp} \norm{\oN\left(\vp\right) - \vt},
            \end{equation}
            where the function $\oN(\vp)$ computes the network's outputs for the corresponding training inputs.
            Such parameters typically represent the biases and weights of thousands of neurons and millions of connections~\cite{Goodfellow2016, Devlin18}.
    		Maxwell's equations can be rewritten in a similar form.
        
            For coherent light with an angular frequency, $\omega$, Maxwell's equations in an inhomogeneous material can be written as the time-independent complex vector functions
            \begin{eqnarray}
                \nabla \times \vE(\pos) \;\;\; = & \quad\;\; - \frac{\partial \vB(\pos)}{\partial t} & =  \qquad\quad \ii \omega \vB(\pos),\label{eq: 3rd Maxwells}\\
                \nabla \times \vH(\pos) \;\;\; = & \vj(\pos) + \frac{\partial \vD(\pos)}{\partial t} & = \;\;\;\; \vj(\pos) - \ii \omega \vD(\pos).\label{eq: 4th Maxwells}
            \end{eqnarray}
            
            \noindent We aim to determine the distribution of the electric field, $\vE(\pos)$, caused by a source of electromagnetic radiation, $\vj(\pos)$. Other quantities such as the electric displacement vector field, $\vD(\pos)$, the magnetic flux density, $\vB(\pos)$, and the magnetising field, $\vH(\pos)$, follow algebraically from $\vE(\pos)$. Without loss of generality, we consider the constituent relations $\vD(\pos) = \epsilon_0 \epsilon_r(\pos)\vE(\pos)$ and $\vB(\pos) = \mu_0 \vH(\pos)$, where only the relative permittivity, $\epsilon_r(\pos) = n^2(\pos)$, is a complex function of space. More general relations can account for anisotropy, chiral, and magnetic properties~\cite{Vettenburg19}. In a dielectric, Eq.~\eqref{eq: 3rd Maxwells} and Eq.~\eqref{eq: 4th Maxwells} can be combined into the well-known vector Helmholtz equation
            \begin{equation}
                \nabla^2 \vE(\pos) +
                k_0^2\epsilon_r(\pos) \vE(\pos) = -\ii \omega \mu_0 \vj(\pos), \label{eq:helmholtz}
            \end{equation}
            where $k_0 = \omega\sqrt{\epsilon_0\mu_0}$ is the vacuum wavenumber.
            In principle, this partial differential equation can be solved numerically by discretising space and determining the field, $\vE$, for a source, $\vj$, on a sufficiently dense sampling grid. Omitting the dependency on position, $\pos$, we can write Eq.~\eqref{eq:helmholtz} succinctly in matrix form
            \begin{equation}
                \oM \vE = \vj, \quad \textrm{ with } \oM \defeq \frac{\ii}{\omega \mu_0} \left(\nabla^2 + k_0^2\epsilon_r\right).\label{eq:helmholtz_matrix}
            \end{equation}    
            The matrix, $\oM$, is generally too large to be represented directly in computer memory. However, the result of its multiplication with any vector can be computed efficiently. This calculation consists of two terms: a convolution, and a multiplication with the permittivity distribution $k_0^2\epsilon_r$. Although the inversion of $\oM$ is infeasible for all but the smallest problems, it is possible to find a unique solution $\vE = \oM^{-1} \vj$ using the following minimisation problem for the parameters, $\vp$,
            \begin{equation}\label{eq:em_optimisation}
                \vE = \argmin_{\vp} \norm{\oM \vp - \vj}.
            \end{equation}
            The parallels with Eq.~\eqref{eq:neural_network_optimisation} are apparent. The neural network's response, $\oN(\vp)$, is replaced by the electromagnetic response, $\oM$, while the training data, $\vt$, is substituted by the current-density, $\vj$, the source of the electromagnetic radiation. There are however notable differences. The neural network function, $\oN(\vp)$, is generally real and non-linear, though machine learning applications do not require a strict globally-optimal solution. In contrast, Maxwell's equations are complex and linear, where an accurate solution is essential. It is important that the minimisation of Eq.~\eqref{eq:em_optimisation} converges deterministically to the true solution.
            
            A one-to-one mapping can be found between Helmholtz Eq.~\eqref{eq:helmholtz} and the neural network with parallel convolutional and multiplication layers depicted in Fig.~\ref{fig:intro}a. The activity of each input neuron on the left corresponds to a specific polarisation of the electric field, $\vE(\pos)$, at each spatial position, $\pos$. Similarly, the activity of the corresponding output neurons on the right represent the source, $\vj(\pos)$, that produced the electric field. I.e.~this particular network \emph{infers} the source, $\vj$, that produces a given field, $\vE$. Generally the significantly more challenging reverse operation is required. Calculating the scattered light field, $\vE$, is equivalent to training the neural network until its output matches the target source, $\vj$. This can be achieved using standard deep learning training algorithms such as stochastic gradient descent (SGD) and adaptive moment estimation (Adam)\cite{Kingma14}. While these algorithms eventually converged for a system of $16 \times 16$ wavelengths, we found that the training was impractically inefficient for all but the smallest problems.
            
            To accelerate the training, we defined an equivalent neural network with an additional hidden layer (Fig.~\ref{fig:intro}b). This network is derived by left-multiplying both sides of Eq.~\eqref{eq:helmholtz_matrix} by $\Gamma^{-1}$, the inverse of the non-singular preconditioner recently proposed by \citeauthor{Osnabrugge16}\cite{Osnabrugge16} and extended to electromagnetism~\cite{Krueger17,Vettenburg22}, so that we obtain
            \begin{equation}
                \preinv\oM\vE = \preinv\vj \quad\textrm{where}\quad \Gamma \defeq \left(s\oV - \oM\right)\oV^{-1}\label{eq:prec_equality}.
            \end{equation}
            The modified potential, $\oV \defeq -\id - k_0^2 \left(\epsilon_r - \epsilon_0\right) / \left(\ii \omega\mu_0 s\right)$, is a representation of the optical properties of the material. The complex scaling constant, $s$, must be chosen so that $\norm{\id + \oV} < 1$ and $\Re{\inp{\vp, s^{-1}\oM\vp}} \ge 0$ for all $\vp$. This is always possible for gain-free systems~\cite{Vettenburg22,Vettenburg22}. For optimal convergence, the background permittivity $\epsilon_0 \in \bbC$ can be chosen to minimise $\max_{\pos}\left|\epsilon_r(\pos) - \epsilon_0\right|$.  
            The inverse preconditioner is written more conveniently as $\Gamma^{-1} = -s^{-1}\oV\Green$, using the spatially-invariant Green's function, $\Green \defeq \left(s^{-1}\oM - \oV\right)^{-1}$. This function can be implemented as a convolutional neural network layer that performs the inversion $\Green \defeq \cF^{-1}\left[\frac{1}{\ii\omega \mu_0 s}\left(\norm{\vk}^2 - k_0^2\epsilon_0\right) + 1\right]^{-1}\cF$, where $\cF$ denotes the Fourier transform. Eq.~\eqref{eq:prec_equality} can now be rewritten as $\oV\left(\id + \Green\oV\right)\vE = -\Gamma^{-1}\vj$, and solved by training the neural network in Fig.~\ref{fig:intro}b.            
            Although, the modified network of Fig.~\ref{fig:intro}b has an extra layer compared to that of Fig.~\ref{fig:intro}a, we found that the use of preconditioning led to an $8$-fold reduction in training time.
            
            The training phase can be completely eliminated by using a recurrent network. After preconditioning, the solution to Eq.~\eqref{eq:prec_equality}, and therefore Eq.~\eqref{eq:helmholtz_matrix}, can be written as the Neumann series $\vE = \sum_{i=0}^\infty \left(\id - \Gamma^{-1}M\right)^i \Gamma^{-1}\vj$. The resulting recurrence relation $\vE_{i+1} = \vE_{i} + \oV\vE_{i} + \oV \Green \oV \vE_{i} + \oV \Green \frac{-1}{s}\vj$ translates into the recurrent neural network depicted in Fig.~\ref{fig:intro}c. In contrast to the former two networks, the recurrent network takes the source, $\vj$, as input and produces the electric field, $\vE$, at its output layer. Since all connection weights are predefined by the material properties and the physics, this network does not require any training. It is sufficient to present the light source as a complex function, $\vj(\pos)$, to infer the a priori unknown electric field distribution, $\vE(\pos)$. While the Scattering Network does not require machine \emph{learning} in the literal sense of the word, its topology unlocks the potential of machine learning and deep learning frameworks for wave scattering computations. Although the accuracy of the solution can be guaranteed mathematically~\cite{Vettenburg22}, it can also be verified using Helmholtz equation~\eqref{eq:helmholtz}, modelled by the network of the forward problem (Fig.~\ref{fig:intro}a).
        
            We implemented the Scattering Network method with the machine-learning library PyTorch and integrated it into the open-source electromagnetic solver MacroMax~\cite{MacroMax22}. This provides immediate access to more general material properties, including birefringent and magnetic materials, while the previously CPU-bound calculations can now be seamlessly executed on the latest machine-learning infrastructure. More specifically, we used the publicly-accessible Google Colab to directly study scattering on a scale relevant to microscopy.
    
        \subsection{Coherent optical scattering on the millimetre scale.\hfill}
            \begin{figure*}[t!]
                \centering\includegraphics[width=\textwidth]{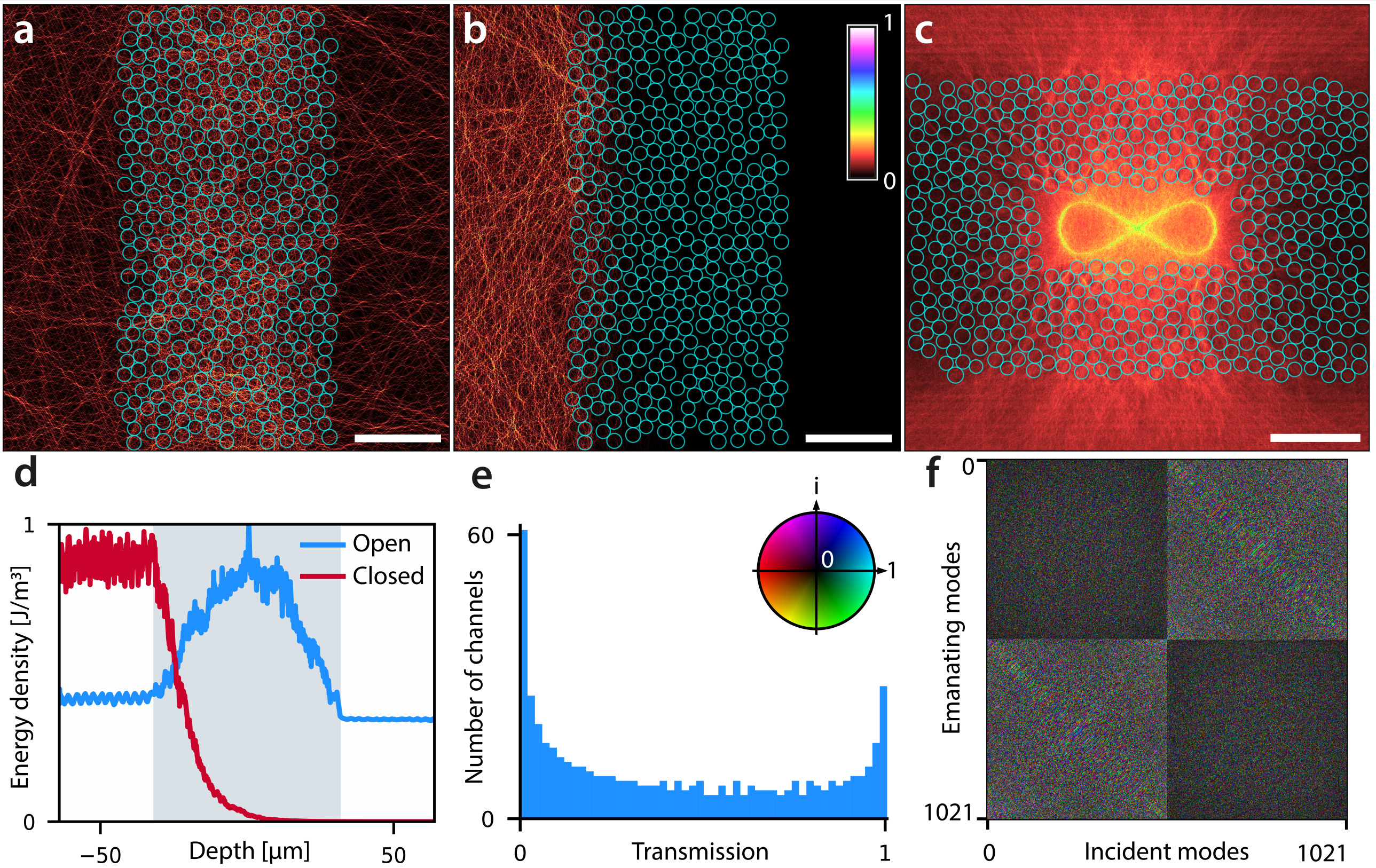} 
                \internallinenumbers
                \caption{\textbf{Transmission through and deposition within a scattering medium}. (\textbf{a}) The intensity distribution of the maximal left-to-right transmission, the open channel. Colourbar in panel (b). All scalebars are $\SI{25}{\micro\meter}$. (\textbf{b}) Intensity distribution of the minimally transmissive mode or closed channel. (\textbf{c}) Deposition of light on a $\infty$-shaped trajectory within a scattering medium. The complete deposition matrix, with all internal fields, had to be calculated to allow control of the field inside the scatterer. (\textbf{d}) Comparison of the energy density in the open and closed channels as a function of propagation depth. (\textbf{e}) Histogram of the transmitted amplitude fraction of all the modes, exhibiting the well-known bimodal distribution~\cite{Rotter17}. (\textbf{f}) Complete scattering matrix with $1022$ modes in total. Complex values are depicted using the colour legend in panel (e). The diagonal quadrants (top left and bottom right) represented transmission in the forward and backwards direction; however, most light is backscattered as can be seen from the brightness of the off-diagonal quadrants (bottom left and top right). Anti-diagonal phase-correlations can be observed in the back-scattering as predicted by \citeauthor{Aubry09}~\cite{Aubry09}.}
                \label{fig:sm_structure}
            \end{figure*}
            
            The computational efficiency and minimal memory requirements of the recurrent approach allow us to scale up wave-scattering calculations. To demonstrate its potential, we determine the field emitted by a \emph{guide-star} embedded deep within a multi-millimetre-wide heterogeneous scattering medium, and use it for phase-conjugation refocussing. Fig.~\ref{fig:intro}f shows the emission over a $\SI{5}{\milli\meter} \times \SI{5}{\milli\meter}$-area on both sides of the $\SI{2.5}{\milli\meter}$-thick slab of scattering material. This corresponds to $8,000$ vacuum wavelengths ($\lambda = \SI{633}{\nano\meter}$) in each dimension.
            
            The emanating field is recorded at the left-hand side (Fig.~\ref{fig:intro}f), and its conjugation is used as the source in a second calculation (Fig.~\ref{fig:intro}g). The phase-conjugated wavefront propagates back into the scattering material and can be seen to converge onto the target position (Fig.~\ref{fig:intro}g).
            To evaluate the refocussing capability of the phase conjugation method, we quantified the intensity and sharpness of the focus using the Strehl ratio. The latter is defined as the intensity at the target position relative to that of the equivalent free-space focus. We found that $91.6\%$ of the intensity reached the target volume. Furthermore, the refocussing achieved Strehl ratio of $1.08$. Higher than unity Strehl ratio was enabled by the diffusion in the scatterer. Compared to free-space phase-conjugation, the scatterer allowed the light to reach the refocussing target from more angles and hence improved our effective numerical aperture.
            
            The field within the complete $\SI{5}{\milli\meter} \times \SI{5}{\milli\meter}$ system was calculated to a residue of $0.0001$ on the publicly available Google Colab~\cite{Carneiro18}. The residue, i.e.~the relative error between the left and right-hand sides of Eq.~\eqref{eq:prec_equality}, decreases monotonically~\cite{Vettenburg19, Vettenburg22}.
            While the residue of this exceptionally large system was reduced to $0.0001$ after $39,002$ recurrences ($\SI{108}{\minute}$), a residue of $0.0005$ can already be reached for this $\SI{5}{\milli\meter}$-wide sample after $11,166$ recurrences in $30$ minutes.
            
            At least $\SI{4.3}{\gibi\byte}$ of memory is required to store a single result. When including working memory, its computation requires approximately $8$ times as much, $\SI{35}{\gibi\byte}$. Perhaps more so than floating point operations per second, the availability of high-bandwidth memory determines the size of the largest problems that can be addressed.
            While solving inverse problems with vectors of this size is already challenging, the numerical dispersion associated with finite difference methods typically requires an order of magnitude denser sampling~\cite{Taflove05}. By relying on Fourier space calculations of the Laplacian~\cite{Osnabrugge16}, numerical dispersion is avoided. 
            In contrast to methods such as GMRES~\cite{Saad1986}, the fixed memory requirements of the Scattering Network enable optical scattering calculations on the millimetre scale.

        \subsection{Complete control of scattering.}
            The ability to control coherent light waves in scattering materials has important implications for optical imaging and manipulation in biological samples. Since the scattering in such tissues tends to be highly anisotropic (anisotropy factor $g\approx 0.9$), the transport mean free path can span hundreds of wavelengths~\cite{Jacques13}.
            Studies of optical scattering rely on a combination of theoretical predictions and experimental results~\cite{Gigan22, Bertolotti22, Jauregui22, Bender22b}. While wavefront shaping techniques can control the light field at the opposing side of a scattering medium, the behaviour of the field within the medium is often most interesting~\cite{Bender22}. Analytical solutions necessarily involve approximations that may or may not hold in real-world experiments. On the other hand, laboratory experiments are limited by noise and our inability to capture all light modes in an experimental setup~\cite{Yu16}. Numerical field calculations have the potential to bridge the gap between theoretical predictions and experiments. Albeit memory hungry, computational methods have been developed specifically to compute scattering matrices~\cite{Lin22}. A scattering matrix holds the complete information on how incident light waves are scattered by a complex structure~\cite{Rotter17}; however, not the internal fields. To also control the internal fields, an orders-of-magnitude larger deposition matrix is required~\cite{Bender22}. While systems with tens of modes have been analysed~\cite{Bender22}, here we show that the Scattering Network enables the calculation of large deposition matrices with over $1000$ independent modes.
            
            While it is often sufficient to determine a single transmission matrix to study open and closed channels~\cite{Popoff10, Yu13, Yu15}, here we compute the complete $1022 \times 1022$ scattering matrix. Its four quadrants correspond to the forward and backwards transmission matrices on the diagonal and the two off-diagonal reflection matrices. Its structure is depicted in Fig.~\ref{fig:sm_structure}f, where hue encodes the complex argument of its values and lightness encodes their amplitude. The columns correspond to all plane waves that are incident on the scattering structure from independent directions. Likewise, its rows correspond to the emanating plane waves. Although the scattering material is highly anisotropic, its thickness is larger than the transport mean free path ($1.2\;l_0$). This results in visibly higher values in the off-diagonal reflection quadrants. The diagonals of the reflection matrices also show a peculiar anti-diagonal phase correlation in the anti-diagonal direction. As \citeauthor{Aubry09} noted for acoustic waves~\cite{Aubry09}, this indicates a regime where multiple scattering does not dominate. Fig.~\ref{fig:sm_structure}e shows the distribution of the singular values of the left-to-right transmission matrix (bottom right quadrant). The bi-modal distribution indicates the presence of both open and closed channels~\cite{Popoff10}. Fig.~\ref{fig:sm_structure}a and Fig.~\ref{fig:sm_structure}b show the intensity distribution within a maximally transmissive (open) mode and a maximally reflective (closed) mode, respectively. Although the net energy transfer of the open channel must be constant throughout the sample, the energy density of multiple-scattering peaks at the centre of the slab, as can be seen in Fig.~\ref{fig:sm_structure}d. In contrast, the energy density of the closed channel can be seen to drops off rapidly with depth.

            While interesting, scattering matrices do not contain information about the internal fields. A deposition matrix is required to control the light field at any position within the scattering material. Here, we compute a $640,000 \times 1022$ deposition matrix (too large to show, data can be accessed following instructions in the data availability section.). Its $1022$ columns correspond to the same incident waves as those of the scattering matrix, though its rows refer to positions within the scattering structure. We calculated all the fields in $42$ minutes on Google Colab. Once the deposition matrix is constructed, standard numerical algorithms can be used to compute the optimal incident waves to produce a desired field in the target region. Here, we use it to trace an $\infty$-shaped path inside the scattering system. Unlike the phase-conjugation method, the light-deposition in the surrounding material can be controlled because the deposition matrix also contains information for areas that are not targeted. Minimising sample exposure is of particular importance when imaging biological tissue.
                   
    \section{Discussion}
        \begin{figure}
            \centering
            \includegraphics[width=0.5\textwidth]{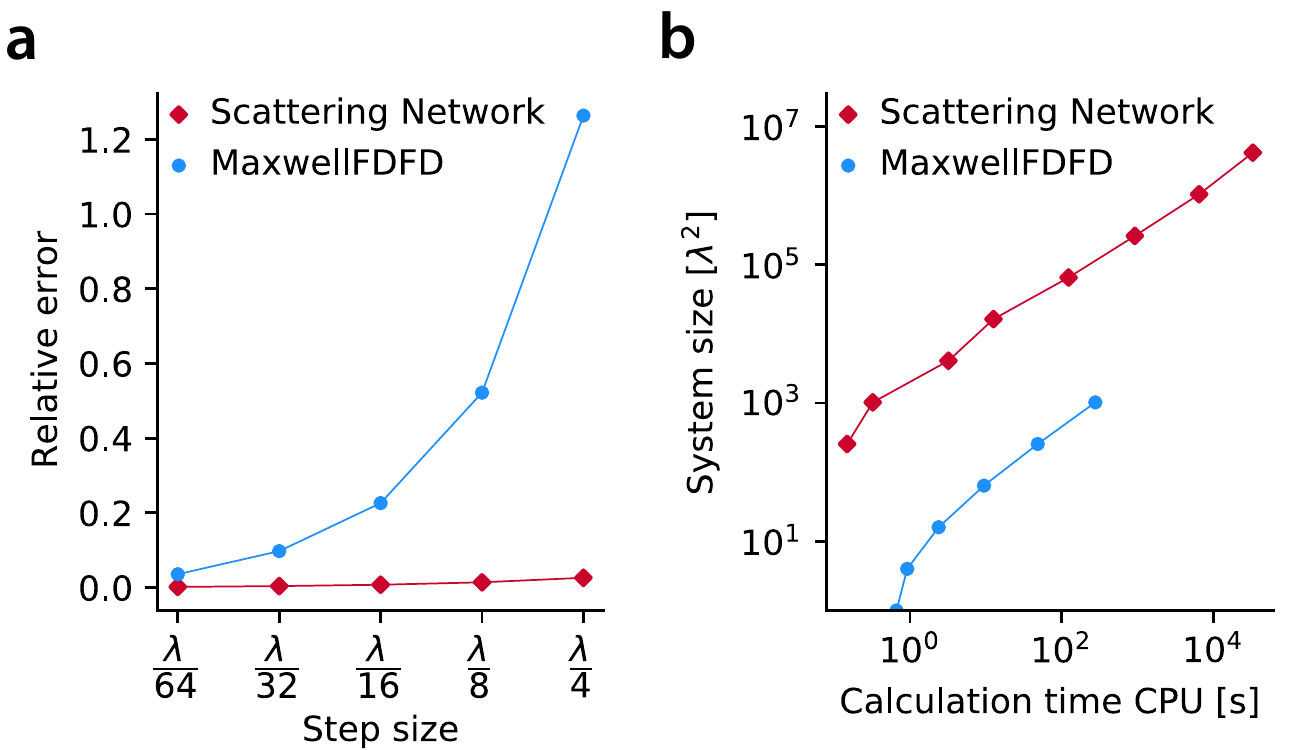}
            \internallinenumbers
            \rightlinenumbers
            \caption{Scaling comparison of the Scattering Network approach with the finite-difference time domain method. (\textbf{a}) Evaluation of the influence of sampling density on accuracy. As the Scattering Network approach uses a Fourier-space convolution instead of finite differences, it avoids numerical dispersion~\cite{Taflove05}. At $64$ samples per wavelength, the sampling error of the Scattering Network is only $0.16\%$. Even at $4$ samples per wavelength, the Scattering Network is more accurate than the finite difference method at $64$ samples per wavelength, resp.~$2.6\%$ vs.~$3.6\%$. This permits coarser sampling without loss of accuracy. (\textbf{b}) The maximum problem size that can be computed within a given time frame. Comparisons are shown for the finite-difference implementation MaxwellFDFD~\cite{Shin12,maxwellfdfd-webpage}, and the Scattering Network integration into MacroMax~\cite{Vettenburg19}, executed on the same CPU and on the cloud. The aforementioned sampling densities of $64$ and $4$ samples per wavelengths are used to ensure comparable accuracy. Systems larger than $1000\,\lambda^2$ could not be calculated using the FDTD approach due to memory limitations ($\SI{64}{\gibi\byte}$).}
            \label{fig:FDFD_vs_Macromax}
        \end{figure}
        
        The topology of the Scattering Network is derived directly from the laws of physics. This is reminiscent of \emph{physics-inspired} or \emph{physics-driven} deep learning methods which employ physical laws in the loss-function to train a neural network. E.g.~this allowed \citeauthor{Lim22} to train a network to calculate light scattering in two-dimensional volumes of approximately $15 \times \SI{20}{\micro\meter}$ and use it for the inverse design of microlenses~\cite{Lim22}. However impressive, the method cannot solve general problems such as those with resonances~\cite{Lim22}. This excludes, for instance, the study of random lasers~\cite{Gottardo08, Cao05, Sapienza22}. A recurrent neural network can also be trained to model the time-dynamics of resonances at concrete points~\cite{Tang22}. Instead of using a physics-based loss-function, the topology of the Scattering Network itself directly reflect the laws of physics. This eliminates the training phase, and with it, the unavoidable training bias. The complete field distributions can thus be calculated with confidence, irrespective of the presence of resonances.
        
        The structure of the Scattering Network lends itself well to machine learning frameworks as PyTorch. This allowed us to capitalise on the rapid developments in the deep learning community and it facilitates deployment on the latest hardware and platforms. As can be seen from the timing comparisons in Figs.~\ref{fig:intro}d and e, enabling cloud-based calculations improves its efficiency by two orders of magnitude. Since the time complexity scales approximately linearly with the problem size, this allowed us to analyse $280$-fold larger systems than those that we could consider up until now. The efficiency advantage is already visible for $1.6 \times 10^4$ sample points and it grows approximately linearly with the problem size. Although this analysis was performed in two dimensions, we found the recurrence time to be insensitive to the number of dimensions.
        
        To fit the largest problems in working memory it is essential that oversampling is avoided. When compared to commonly-used algorithms such as the finite-difference frequency domain method (FDFD), the neural network uses Fourier-space convolutions instead of finite-differences to determine the Laplacian or Green's function. As can be seen from Fig.~\ref{fig:FDFD_vs_Macromax}a, this makes it significantly more accurate, in particular on coarse grids.
        
        The ability to reduce the sampling to a minimum has a direct impact on the size of systems that can be computed. E.g.~when sampled at $\lambda/3$ for $\lambda = \SI{586}{\nano\meter}$, the electromagnetic field of a $\SI{100}{\micro\meter} \times \SI{100}{\micro\meter} \times \SI{100}{\micro\meter}$-volume can be calculated within a working memory of $\SI{24}{\gibi\byte}$. This factors in polarisation, as well as the seven vectors of temporary storage for the problem definition and the temporary memory of our current implementation. Contrast this to a more typical FDFD sampling of $\lambda/30$, which would take as much as $\SI{1}{\tebi\byte}$ just to store the solution for a single polarisation. The Scattering Network Method's limited memory requirements make it ideal for large heterogeneous problems.
        
        The number of sample points, $576$~million for the $\SI{5}{\milli\meter}$-wide system, also plays an important role in the computation time as can be seen from Fig.~\ref{fig:intro}d. However, we found that the number of dimensions has a negligible impact on the recurrence time. For instance, a three-dimensional system spanning $\SI{176}{\micro\meter} \times \SI{176}{\micro\meter} \times \SI{176}{\micro\meter}$ with this number of sampling points requires approximately the same time per recurrence.

        The deterministic response of the Scattering Network is cemented in its one-to-one correspondence with physical laws. By directly mirroring the laws of electromagnetism, the Scattering Network enabled us to leverage powerful, yet publicly available, machine learning infrastructure. This need not be limited to electromagnetism alone. As the convergent Born series was recently extended to a general class of numerical problems~\cite{Vettenburg22}, we anticipate that other problems of science and engineering can be addressed by identifying suitable neural network topologies.
        
        In conclusion, we demonstrated how Maxwell's equations can be rephrased as a recurrent neural network. The recurrent neural network has a single hidden layer which performs a shifted Green's function, while the other connection weights have a one-to-one correspondence to the problem's permittivity distribution. In sharp contrast to conventional machine learning approaches, this avoids the need for training and the associated training bias. By incorporating preconditioning from the outset, convergence is efficient and fully deterministic. Combined with the ability to avoid numerical dispersion, this makes the method particularly attractive to study anisotropic scattering in biological tissue on scales relevant to microscopy.
        
    \methods
        \subsection{Implementation.}
            The electromagnetic solver was implemented using the PyTorch (1.12.1+cu113) machine learning library~\cite{Paszke2019}, and integrated into the MacroMax library~\cite{Vettenburg19}. The widely used PyTorch framework was chosen for its built-in ability to handle complex numbers and perform fast Fourier transforms. Its ubiquity grants access to the latest cloud computing platforms and technological advances. PyTorch's Adam optimiser was used to train the forward neural networks depicted in Fig.~\ref{fig:intro}a-b. We found the Stochastic Gradient Descent algorithm to be less efficient for this network topology. A fast-Fourier-transform-based convolution layer was used to avoid finite differences and calculate the Laplacian more efficiently than with the built-in convolution operation.
            The recurrent neural network of Fig.~\ref{fig:intro}c also implements the Green's function as such convolution. Unlike the preceding topologies, the network connections are pre-determined by the physics of the problem and its output layer directly presents the solution. As such, this network does not require training and it can infer the solution to large problems with minimal memory requirements. We integrated this approach with the electromagnetic solver, MacroMax~\cite{MacroMax22}, to make it readily accessible and extend its use to e.g.~birefringent and chiral materials.
        
        \subsection{Scattering system parameters.}
            The $\SI{5}{\milli\meter} \times \SI{5}{\milli\meter}$ scattering medium shown in Fig.~\ref{fig:intro}f-g consists of packed spheres with a radius of $\SI{30}{\micro\meter} \pm 10\,\%$ and a refractive index of $1.33$. The scattering layer spans the entire width and is $\SI{2.5}{\milli\meter}$ thick. In this example, the material structure and fields were sampled at $\lambda/3$ using a vacuum wavelength of $\SI{633}{\nano\meter}$. The system is padded with absorbing boundaries that are $50$ wavelengths thick and have a linearly increasing extinction coefficient from $0$ to $0.2$.
            
            The scattering and deposition matrix calculations in Fig.~\ref{fig:sm_structure}a-c assume a $\SI{500}{\nano\meter}$ light source. Both systems contain a scattering slab made of spheres with radius $\SI{2.0}{\micro\meter} \pm 5\,\%$ and refractive index $1.33$, centred in a calculation volume of $\SI{128}{\micro\meter} \times \SI{128}{\micro\meter}$ ($256\lambda \times 256\lambda$). The slab thickness for the scattering matrix calculation in Fig.~\ref{fig:sm_structure}(a,b,d-f) is $\SI{64}{\micro\meter}$, while that for the deposition matrix calculation (Fig.~\ref{fig:sm_structure}c) is $\SI{85}{\micro\meter}$ thick. The latter has a $\SI{17}{\micro\meter} \times \SI{51}{\micro\meter}$-deep internal gap at its centre to study the targeted deposition of light.
            
        \subsection{Calculation of deposition and scattering matrices.}
        	We computed the internally and externally scattered fields for a complete basis of incident plane waves. To study transmission matrices we considered a slab geometry, orthogonal to the $z$-axis. To minimise edge effects when simulating an infinitely-wide slab, we adopt periodic boundary conditions in the transverse dimension. Incident and emanating waves on both sides are all represented in a common plane-wave basis with equal irradiance along the $z$-axis. The scattering matrix for free-space propagation would thus be the identity matrix. The basis vectors are listed in raster-scan order for the forward and backwards propagating waves, respectively. In the code accompanying this manuscript~\cite{MacroMax22}, vector waves are represented as a pairs of orthogonal propagating polarisations per plane wave. The scattering and deposition matrices used to produce the examples in Fig.~\ref{fig:sm_structure} each have $1022$ columns, one per independent mode.
            
            The deposition matrix used for Fig.~\ref{fig:sm_structure}c has $640,000$ rows, one for each internal and external sample-point of the field. While the scattering matrix is a square matrix with the same plane wave basis for the row space.
			The scattering matrix can be considered a $2 \times 2$-block matrix with four quadrants. The two quadrants on the diagonal correspond to the forward and backward transmission matrices, while the off-diagonal quadrants are the front and back-reflection matrices~\cite{Rotter17}.
			
            To determine the open and closed channels depicted in Fig.~\ref{fig:sm_structure}a-b, we calculated the singular value decomposition of the forward transmission matrix, (i.e.~top-left quadrant of scattering matrix in Fig.~\ref{fig:sm_structure}f). This was implemented using SciPy, which relies on LAPACK's gesdd divide-and-conquer algorithm.
            To refocus within the deposition region inside the scattering material (Fig.~\ref{fig:sm_structure}c), we calculated the Moore-Penrose pseudoinverse of the deposition matrix. As a first step, the singular value decomposition of the deposition matrix was determined using ARPACK's implicitly restarted Lanczos algorithm. To limit the amplification of numerical errors and the demands on computer memory, the pseudoinverse was calculated using the $750$ dominant vectors. The incident fields that focus to specific points or patterns deep within the scatterer can then be determined from the corresponding columns of the pseudo-inverse.
            
	    \subsection{Time efficiency.}
	        To avoid giving disproportionate weight to the initial recurrence, Figures~\ref{fig:intro}d-e compare the median recurrence time. While the initialisation overhead is negligible for typical problems, it does hamper a direct comparison for the smallest problem sizes. Typical systems require around $1000$ recurrences to converge, depending on the range of permittivity, the size, and structure of the scattering system.
	        The timings shown compare scalar calculations for a range of system sizes (expressed in number of sample points). Vectorial calculations were found to take approximately $4$ times as long.
	        While the time per recurrence grows steadily with the number of sample points in the calculation space, the number of dimensions was found to have negligible influence on the recurrence time.
    
    \bmsection{Data availability}\label{sec:data_availability}
        All data underlying the results was generated by the algorithms and code described in this manuscript. The complex-valued scattering matrix and data shown in Fig.~\ref{fig:intro}-\ref{fig:sm_structure} can be accessed using the following link \url{https://bit.ly/3uxFhZa}. A DOI link will be provided with the published manuscript.
        
    \bmsection{Code availability}
        The algorithm, as well as the data visualisation, is implemented in Python using PyTorch. The complete source code with examples is openly available as a Git repository~\cite{MacroMax22}. The PyTorch implementation is integrated in the MacroMax electromagnetic calculation library, which is freely available on the Python Package Index~\cite{PyPI-MacroMax}.
    
    \printbibliography
    
    \bmsection{Acknowledgements}
        L.~Valantinas' research is supported by EPSRC Grant: EP/R513192/1. T.~Vettenburg is a UKRI Future Leaders Fellow supported by grant MR/S034900/1. For the purpose of open access, the author(s) has applied a Creative Commons Attribution (CC BY) licence to any Author Accepted Manuscript version arising.
        
    \bmsection{Author contributions}
        LV integrated PyTorch into the MacroMax library, performed the numerical calculations, and created all figures. TV developed the theory and a proof-of-principle implementation, he adapted the MacroMax library for multi-platform execution and extended it to automate scattering matrix calculations. Both authors wrote and reviewed the manuscript.
    
    \bmsection{Competing interests}
        The authors declare no competing interests.
        
    \bmsection{Additional information}
        \textbf{Correspondence and requests for materials} can be addressed to\\ \href{mailto:t.vettenburg@dundee.ac.uk}{T.~Vettenburg <t.vettenburg@dundee.ac.uk>}.

\end{document}